# Personalized Detection of Stress via hdrEEG: Linking Neuro-markers to Cortisol, HRV, and Self-Report

**N. B. Maimon[1,2], Ganit Baruchin[3,4], Itamar Grotto[3], Nathan Intrator[2,5,6], Lior Molcho[2], Talya Zeimer[2], Ofir Chibotero[2], Nardeen Murad[7], Yori Gidron[7], and Efrat Danino[4]**

[1] School of Psychological Sciences, Tel Aviv University, Tel Aviv, Israel

[2] Neurosteer Inc, NYC, New York, USA

[3] School of Public Health, Ben-Gurion University of the Negev, Beer Sheva, Israel

[4] Shamir Academic School of Nursing in the Hebrew University in Jerusalem, Israel

[5] Blavatnik School of Computer Science, Tel-Aviv University, Tel-Aviv, Israel

[6] Sagol School of Neuroscience, Tel Aviv University, Tel Aviv, Israel

[7] Department of Nursing, Faculty of Social Welfare and Health Sciences University of Haifa, Haifa, Israel

**Abstract**

Chronic stress is a major risk factor for cognitive decline and systemic illness, yet reliable individual-level neural markers remain limited. We tested whether two single-channel high-dynamic-range EEG (hdrEEG) biomarkers, ST4 and T2, index personalized stress responses by linking neural activity to validated physiological and subjective measures. Two studies were conducted. Study 1 included 101 healthy adults (22–82 years) who completed resilience, burnout, and stress questionnaires, provided salivary cortisol, and performed resting, low-load, high-load, emotional, and startle conditions during hdrEEG. Study 2 included 82 healthy adults (19–42 years) who completed the State-Trait Anxiety Inventory, underwent heart-rate-variability (HRV) monitoring, and performed auditory, stress-inducing, and emotional conditions during hdrEEG. To balance exploration and control of false discovery, correlations were interpreted as meaningful at $r >= 0.30$, with p-values reported.

ST4 reflected physiological arousal and cognitive strain. In Study 1, resting ST4 correlated positively with cortisol ($r = 0.54$, $p < 0.001$) and was lower among more resilient participants during a low-load attention condition ($r = -0.41$, $p = 0.026$). In Study 2, ST4 correlated negatively with HRV: RMSSD during anticipatory stress and post-stress recovery ($r = -0.24$, $p = 0.028$; $r = -0.32$, $p = 0.04$) and SDNN during recovery and cognitively demanding conditions ($r = -0.41$, $p = 0.01$; $r = -0.47$, $p = 0.003$). T2 captured emotional and autonomic regulation. In Study 1, greater T2 differences between high mental load and rest tracked higher cortisol ($r = 0.32$, $p = 0.038$), T2 during emotionally charged words was lower with higher resilience ($r = 0.31$ for both, $p < 0.05$). In Study 2, T2 was higher with greater trait anxiety ($r = 0.367$, $p = 0.026$) and correlated negatively with SDNN during stress-related and emotional conditions ($r = -0.33$, $p = 0.029$; $r = -0.35$, $p = 0.021$) and with RMSSD during emotional conditions ($r = -0.32$, $p = 0.036$).

Together, ST4 and T2 provide complementary, portable hdrEEG markers of the human stress response—ST4 indexing physiological arousal and cognitive strain, T2 indexing emotional-autonomic regulation—supporting individualized assessment in clinical and real-world contexts.

## 1. Introduction

Individual differences in stress reactivity and regulation represent a critical challenge for objective neurophysiological assessment, given the profound implications of chronic stress on health outcomes, cognitive function, and brain health [1]. Chronic stress exposure has been linked to accelerated cognitive decline, increased risk of neurodegenerative diseases, and systemic inflammation that may contribute to Alzheimer's disease pathogenesis [2]. The hypothalamic-pituitary-adrenal (HPA) axis dysregulation associated with chronic stress promotes neuroinflammation through sustained cortisol release and glucocorticoid resistance, creating a cascade of inflammatory processes involving IL-1β, IL-6, and TNF-α that directly impact brain regions including the hippocampus [3]. Moreover, elevated cortisol levels predict hippocampal atrophy, memory decline, and increased Alzheimer's disease risk, underscoring the critical need for reliable individual-level stress biomarkers [4,5].

### 1.1 Physiological and Subjective Stress Biomarkers

#### *1.1.1. Cortisol Measurement*

Salivary cortisol has emerged as the gold standard biomarker for HPA axis function and stress assessment, reflecting free cortisol levels that correlate strongly with serum concentrations [5]. The diurnal cortisol rhythm, characterized by the cortisol awakening response (CAR) and diurnal slope, provides critical information about HPA axis regulation [5]. Disrupted cortisol patterns, including flattened awakening responses and abnormal diurnal slopes, serve as consistent markers of HPA axis dysfunction and are associated with various health outcomes including psychiatric illness, cardiovascular mortality, and cognitive decline [3,5].

#### *1.1.2. Heart Rate Variability*

Heart rate variability (HRV) represents the variation in time between successive heartbeats and reflects autonomic nervous system balance between sympathetic and parasympathetic activity [6]. High HRV indicates intact autonomic adaptability and parasympathetic dominance, while reduced HRV reflects sympathetic hyperactivation associated with stress and illness [7]. HRV measures provide insight into stress reactivity through the body's physiological response to cognitive and emotional demands, though individual differences in stress response patterns can affect interpretation [8].

#### *1.1.3. Subjective Measures*

The State-Trait Anxiety Inventory (STAI) represents the most widely used validated instrument for assessing both transient state anxiety and stable trait anxiety predisposition [9]. The STAI's 40-item structure (20 items each for state and trait anxiety) on 4-point Likert scales has demonstrated high internal consistency ($\alpha$ = .86-.95) and test-retest reliability (.65-.75) across diverse populations [9]. This instrument effectively captures individual differences in anxiety proneness and immediate stress responses, providing crucial subjective context for neurophysiological measures [9,10].

#### *1.1.4. EEG-Based Individual Stress Detection*

Recent advances in EEG-based stress detection have demonstrated promising individual-level classification accuracies, though most studies focus on group-level rather than personalized biomarkers [11,12]. Multi-modal approaches combining EEG with peripheral physiological signals have achieved 98.1% accuracy in stress type classification and 97.8% in stress level detection using 4-second neurophysiological signals, with EEG outperforming peripheral measures for longer recording periods [13]. Single-channel EEG systems have shown particular promise for portable stress monitoring, with studies reporting 81.0% accuracy using dry electrode headbands during cognitive stress tasks [14,15].

Frequency-domain EEG analyses consistently reveal stress-related changes in spectral power, particularly increased frontal gamma and beta activity during acute stress, alongside decreased alpha power reflecting cortical activation [13,14]. Theta band increases correlate with cognitive load rather than stress per se, suggesting dissociable neural signatures for different psychological states [15]. However, individual differences in stress sensitivity and EEG response patterns remain poorly characterized [11,12].

Machine learning approaches have enhanced stress detection capabilities, with studies demonstrating that personalized models accounting for individual response patterns significantly outperform generalized classifiers [11,15]. Low-cost consumer EEG devices combined with machine learning methods provide viable alternatives to medical-grade systems, though standardization of signal processing and sensor placement remains challenging [16].

### 1.2 Research Gap and Innovation

Despite extensive research on EEG stress biomarkers, no current assessment protocol combines a mobile

EEG device with specifically designed cognitive tasks that demonstrate stress biomarkers correlating with established physiological and subjective measures at the individual level [17,18]. Most existing studies focus on group-level effects or short-term laboratory stress induction rather than individual differences in trait-like stress reactivity and regulation patterns [17-20].

The Neurosteer high dynamic range EEG (hdrEEG) system offers a novel single-channel approach that has been validated for cognitive load detection and early Parkinson's disease assessment using auditory cognitive tasks [9]. This FDA-cleared system employs prefrontal electrode placement (Fp1-Fp2 differential) with machine learning-derived features that have shown sensitivity to individual differences in cognitive function and decline [18,19]. The addition of emotional stimuli through lexical decision tasks with positive and negative valence words, combined with validated happy and sad music excerpts, provides a comprehensive assessment of emotional reactivity alongside cognitive load [17,20].

The present research addresses this gap by investigating correlations between Neurosteer-derived EEG biomarkers and multiple validation measures including cortisol levels, HRV, and extensive subjective assessments of stress and resilience. This multi-modal validation approach in large-scale studies (>100 and ~80 participants) enables identification of reliable individual-level stress biomarkers that could transform clinical and occupational stress assessment through objective, portable neurophysiological monitoring [17-20].

## 2. Methods

Study 1. Ethical approval for Study 1 was obtained from the Shamir Medical Institute, approval number: ASF-0237-24, 03022025. All participants provided written informed consent prior to participation. The study included 102 healthy adult volunteers (80% women). Data from 101 participants were included in the analyses. Participants ranged in age from 22 to 82 years (M = 48.0). All individuals were screened to ensure they had no history of neurological or psychiatric disorders.

Study 2. Ethical approval for Study 2 (no. 048/23) was granted by the Ethics Committee of Haifa University on Feb 8, 2023. All participants provided written informed consent prior to participation. The study included 82 healthy adult volunteers, of whom 65% were women. Participants ranged in age from 19 to 42 years (M = 28.8). As in Study 1, participants were screened to confirm absence of neurological or psychiatric disorders.

### 2.2. Aparatus

EEG measurements were executed utilizing the Recorder (Neurosteer EEG recorder). An FDA cleared adhesive with three electrodes was applied to the subject’s forehead, using a dry gel to enhance signal quality.

The non-intrusive electrodes were located at the prefrontal areas, producing a single-EEG channel derived from the difference between Fp1 and Fp2 and a ground electrode at Fpz, based on the international 10/20 electrode positioning. The signal range is ±25 mV (background noise <30nVrms). The electrode contact impedances were kept under 12 kΩ, as determined by a handheld impedance device (EZM4A, Grass Instrument Co., USA). The data was acquired in a continuous mode and subsampled to 500 samples per second.

During the data collection, a proficient research member oversaw each subject to reduce potential muscle interference. Subjects received guidance to refrain from making facial gestures during the session, and the supervising member would notify them if noticeable muscle or eye movements were detected. Notably, the differential signal processing and superior common-mode rejection ratio (CMRR) contribute to minimizing motion disturbances and electrical interference.

### 2.3. Procedure

#### 2.3.1. Study 1

The experiment took place in a tipi tent located in the forest, prior to participants' engagement in a forest treatment program. Upon arrival, participants completed a battery of self-report questionnaires assessing stress, resilience, and anxiety:

- Emotion Regulation Questionnaire (ERQ; [21]). The ERQ measures individual differences in two key strategies of emotional regulation: cognitive reappraisal (changing one's interpretation of a situation to alter its emotional impact) and expressive suppression (inhibiting outward signs of emotion). It is widely used to evaluate resilience and adaptive versus maladaptive coping. Items are rated on a Likert scale and yield separate subscale scores.
- Maslach Burnout Inventory (MBI; [22]). The MBI is the gold-standard instrument for assessing burnout, particularly in occupational contexts. It measures three dimensions: emotional exhaustion, depersonalization (detachment from work or people), and reduced personal accomplishment. Higher scores on exhaustion and depersonalization, alongside lower accomplishment, indicate higher levels of burnout.
- Stress Questionnaire for Health Professionals [23]. This questionnaire assesses perceived stress specific to work-related demands, including workload, emotional strain, and interpersonal challenges in professional environments. It is frequently used to capture stress burden in caregiving and high-responsibility occupations, providing a domain-specific measure of acute and chronic stress levels.

After completing the questionnaires, participants provided a saliva sample for cortisol analysis, an objective biomarker of physiological stress response. They then underwent EEG examination in groups of four. Each participant was seated separately to avoid visual contact. The tipi environment was kept quiet and air-conditioned to minimize confounds.

### 2.3.2. Study 2

Participants first completed the State-Trait Anxiety Inventory (STAI; [9]). The STAI consists of two 20-item subscales measuring state anxiety (the temporary, situational experience of anxiety in the moment) and trait anxiety (the stable predisposition to experience anxiety across time and situations). Items are rated on a four-point Likert scale, and the inventory is highly validated for distinguishing situational stress from chronic anxiety tendencies.

Following the questionnaire, participants were connected to a heart rate variability (HRV) recording device and the hdrEEG system. The experiment included two counterbalanced stages:

1. Auditory EEG assessment. Participants listened to auditory stimuli while EEG and HRV responses were recorded.

2. Stress-inducing tasks:

- Job interview simulation: Participants were informed they would perform a mock job interview in front of the experimenter. They were given two minutes to prepare and then spoke for two minutes about why they should be hired.
- Mental arithmetic challenge: Participants were presented with a number and instructed to repeatedly subtract three for one minute, after which they reported the final result.

HRV data were recorded continuously and segmented by task, in parallel with the hdrEEG measurements.

### 2.3.2. EEG recording and auditory assessment protocol

EEG data were recorded using the Neurosteer® single-channel high dynamic range EEG (hdrEEG) Recorder. A three-electrode medical-grade patch was applied to each participant's forehead using dry gel for optimal signal transduction. The monopolar electrode configuration included electrodes positioned at Fp1 and Fp2, based on the International 10/20 system, with a reference electrode at Fpz. The EEG signal was continuously sampled at 500 Hz and transmitted wirelessly for real-time data processing.

EEG signals were processed using time-frequency analysis to extract relevant neural features. Biomarkers That were calculated in this research were ST4 and T2, and were derived using Neurosteer's proprietary machine-learning algorithms. Previous studies have identified ST4 as an indicator of cognitive decline and individual task performance, while T2 differentiate between cognitive load and resting state activity within clinical populations.

### 2.3.3. Cognitive Tasks

This study included a previously described auditory detection task (28), an auditory n-back task, and resting state tasks (see Figure 1).

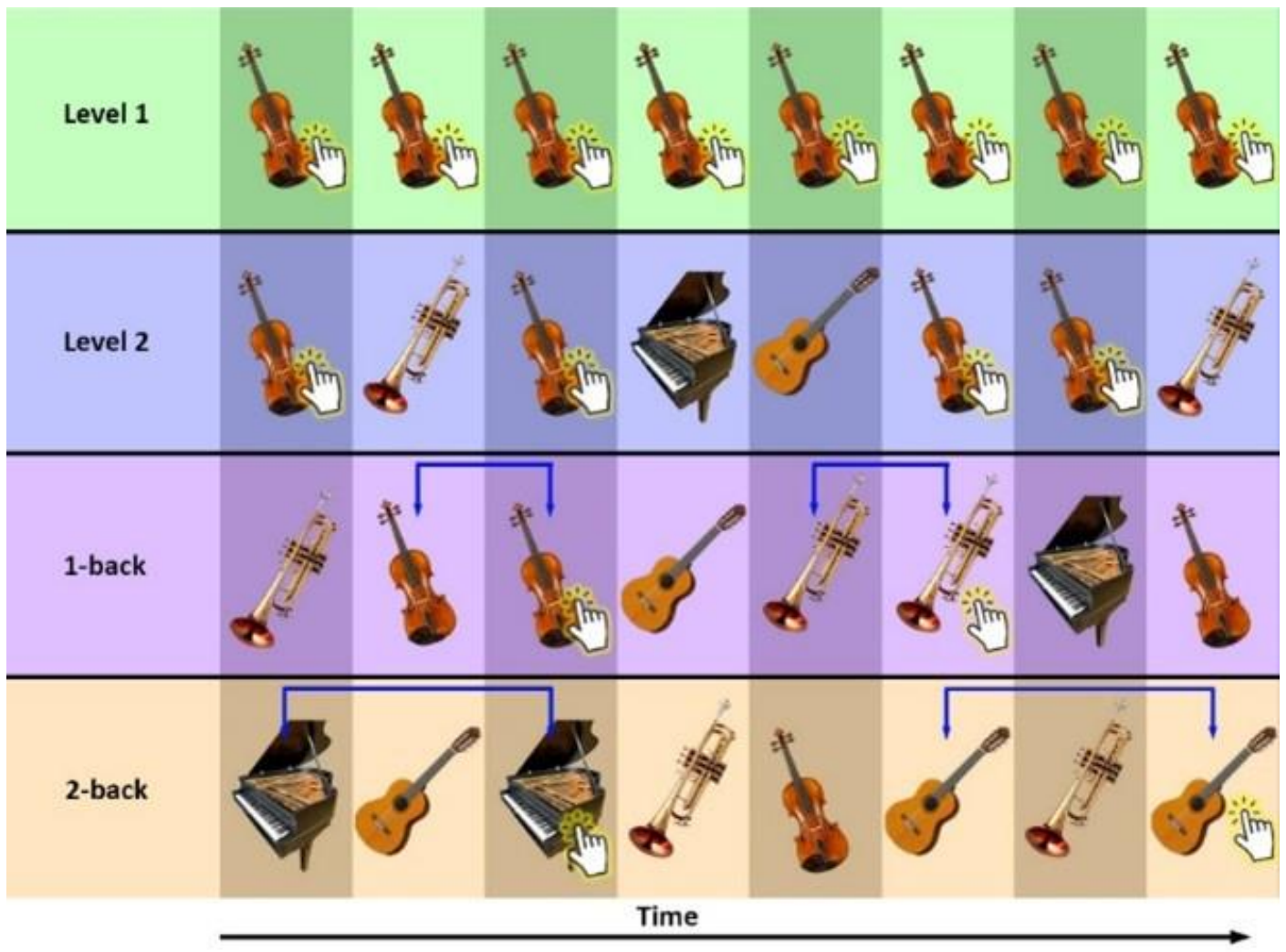

**Figure 1.** A visual representation of the two cognitive tasks used in this study is provided. Auditory Detection: Level 1 features the same melody played by the same musical instrument several times, and the participant is asked to click each time the melody is played. Level 2 presents melodies played by different instruments, and the participant is asked to click only when a melody by a specific instrument is played (in this example, the flute melody). Musical n-back: Levels 1 and 2 showcase melodies played by various instruments. In Level 1, the participant is asked to click whenever a melody is played, while in Level 2, the participant is asked to click only when a melody immediately repeats itself (regardless of which melody is played).

The detection task included a sequence of tunes from a violin, trumpet, and flute. Participants held a clicker in order to respond to the musical cues. Instructions directed participants to click once when they heard a specific instrument playing. Responses were limited to "yes" trials corresponding to the designated instrument's tune. The task was designed with two levels of difficulty to evaluate escalating cognitive demands. In Detection level 1, a consistent tune played for 3 s, recurring throughout the block. Participants were directed to click promptly for every repetition of the tune. This level featured three trials of 90 s each (corresponding to each instrument), with each melody appearing 5–6 times and intervals of 10–18 s of silence. Detection level 2 included tunes lasting for 1.5 s, of three instruments intertwined within a single block. Participants were instructed to respond solely to a designated instrument in that block, disregarding the rest. Each trial in this level had 6–8 melodies interspersed with 8–14 s of silence, and the target tune played 2–3 times.

In the n-back task, participants were presented with a sequence of melodies played by different instruments and used the clicker to respond to the stimuli. This task also included two difficulty levels (0-back and 1-back) to examine increasing cognitive load. A set of melodies (played by a violin, a trumpet, and a flute) was played in a different order for each block, and participants were asked to click a button when the melody repeated n-trials ago (based on the block level). In the 0-back level, participants clicked the button each time a melody was heard. This level included one 90-s block with 9 trials (instances of melody playing), each melody played for 1.5 s and 6–11 s of silence in between. In the 1-back level, participants clicked the button each time a melody repeated itself (n = 1). This level included two 90-s blocks with 12–14 trials (instances of melody playing), each melody played for 1.5 s and 4–6 s of silence in between. In each block, 30–40% of the trials were the target stimulus, where the melody repeated itself, and the participant was expected to click the button.

The resting state condition consisted of three 1-minute trials designed to capture a range of internal states. In the first trial, participants were instructed to keep their eyes open and let their thoughts wander. In the second trial, they were asked to close their eyes and relax, and in the third, they listened passively to a short excerpt of calming meditative music. This design allowed us to record resting neural activity across varying levels of sensory input and cognitive engagement.

To induce acute stress, startle stimuli were presented during one task block per session. These stimuli consisted of lateralized auditory bursts: a ~200 Hz pure sinusoidal tone played for 200 ms with a 50 ms silent gap between bursts. Each burst was delivered either to the left or right ear, with white noise played simultaneously in the opposite ear. The intensity of the tones was approximately 100 dB, and the inter-stimulus interval varied randomly. Acoustic analysis confirmed that both the waveform and frequency spectrum matched these parameters, with a clear peak at ~200 Hz and transient burst patterns.

Finally, to include stimuli that evokes emotional valence, two additional tasks were added:

Lexical Decision Task: Participants completed a block in which they pressed the button only when they detected a real Hebrew word, ignoring non-word stimuli. The presented words carried either positive (e.g., happiness) or negative (e.g., worry) emotional valence. This task allowed assessment of emotion-sensitive lexical processing as participants discriminated between emotionally charged real words and phonologically plausible non-words.

Emotional Music Task. Participants listened to two musical excerpts selected for their well-established emotional valence: Adagio in G Minor (commonly—but erroneously—referred to as “Albinoni’s Adagio”) to evoke sadness, and “All You Need Is Love” by The Beatles to evoke happiness. The Adagio in G Minor has been repeatedly used in prior research to reliably induce sadness in listeners. While the piece's attribution to Albinoni is disputed, its melancholic tone in the minor key reliably elicits negative emotional responses [24]. Although direct empirical validation for “All You Need Is Love” as inducing happiness was not found, its cultural legacy and widespread interpretation as a joyful, uplifting anthem support its use as a happy valence stimulus—especially against the broadly positive sentiment it represents in popular culture and media.

### 2.3.4. Signal Processing

The EEG signal was decomposed into multiple components using harmonic analysis mathematical models [22, 23], and ML methods were employed on the components to extract higher-level EEG features. The full technical specifications for signal processing can be found in Molcho et al. [18,19]. In summary, the Neurosteer® signal-processing algorithm analyzes EEG data using a time/frequency wavelet-packet analysis. This analysis, previously conducted on a separate dataset of EEG recordings, identified an optimal orthogonal basis decomposition from a large collection of wavelet packet atoms, optimized for that set of recordings using the Best Basis algorithm [17-20]. This basis results generated a new representation of 121 optimized components called Brain Activity Features (BAFs). Each BAF consists of time-varying fundamental frequencies and their harmonics.

The BAFs are calculated over a 4-s window, which contains 2,048 time elements due to the 500 Hz sampling frequency. In this window, each BAF is a convolution of a time/frequency wavelet packet atom, allowing for a signal that can vary in frequency over the 4-s window, such as a chirp. The window is then advanced by 1 s, similar to a moving window spectrogram with 75% overlap, and the calculation is repeated for the new 4-s window. The EEG power spectrum is obtained using a fast Fourier transform (FFT) of the EEG signals within a 4-s window.

The data was tested for artifacts due to muscle and eye movement of the prefrontal EEG signals (Fp1, Fp2). The standard methods used to remove non-EEG artifacts were all based on different variants of the Independent Components Analysis (ICA) algorithm [18]. These methods could not be performed here, as only a single-channel EEG data was used. As an alternative, strong muscle artifacts have higher amplitudes than regular EEG signals, mainly in the high frequencies; thus, they are clearly observable in many of the BAFs that are tuned to high frequency. This phenomenon helps in the identification of artifacts in the signal. Minor muscle activity is filtered out by the time/frequency nature of the BAFs and thus caused no disturbance to the processed signal. Similarly, eye movements are detected in specific BAFs and are taken into account during signal processing and data analysis.

### 2.3.5. Construction of Higher-Level Classifiers

Several linear combinations were obtained using ML techniques on labeled datasets previously collected by Neurosteer ® using the described BAFs. Specifically, EEG feature ST4 was calculated using PCA [19]. Principle component analysis is a method used for feature dimensionality reduction before classification. Studies show that features extracted using PCA show significant correlation to MMSE score and distinguish

AD from healthy subjects [18,19], as well as show good performance for the diagnosis of AD using imaging (Choi and Jin, 2018). Here, the fourth principal component was found to separate between low and high difficulty levels of auditory n-back task for healthy participants (ages 30–70). T2 activity extracted during the cognitive auditory assessment was found to correlate with subjective measurements of resilience [17]. Most importantly, these two EEG features were derived from different datasets than the data analyzed in the present study. Therefore, the same weight matrices that were previously found were used to transform the data obtained in the present study.

### 2.4. Statistical Analysis

EEG data were averaged across trials for each participant and categorized into three task conditions: resting state, stress, emotional tasks and mental load. The resting state condition included all resting state tasks, the stress condition included the abrupt auditory beeps and stressful tasks (study 2), the mental load condition combined both the n-back and detection tasks, and emotional condition included the 2 musical excerpts and words in the lexical task. For each participant, neural activity within these categories was averaged to create a single value per feature and condition.

For both studies, analyses focused on the EEG biomarkers ST4 and T2, which were examined in relation to subjective and physiological measures of stress and resilience. All measures were computed separately for each experimental task condition (resting state, mental load, emotional, and stress), and correlations were performed at the participant level.

Study 1: Pearson correlations were computed between ST4 and T2 activity and the self-report measures derived from the Emotion Regulation Questionnaire (ERQ), the Maslach Burnout Inventory (MBI), and the Stress Questionnaire for Health Professionals. For each instrument, mean composite scores were calculated to reflect overall levels of resilience, exhaustion/burnout, and perceived stress, respectively. In addition, correlations were calculated with salivary cortisol levels obtained prior to the EEG recording.

Study 2: Pearson correlations were calculated between ST4 and T2 activity and physiological indices of autonomic regulation, measured via heart rate variability (HRV). HRV features were segmented by task, allowing assessment of biomarker–physiology associations across both the auditory and stress-inducing tasks.

To account for multiple comparisons while maintaining exploratory sensitivity, we applied a threshold of $r \geq 0.30$ to interpret correlations as meaningful, a commonly used criterion in behavioral and neuroscience research to reduce false discovery rates [26].

## 3. Results

### 3.1 Study 1

ST4 activity during resting state was higher among participants who exhibited elevated cortisol levels (R = 0.54, p < 0.001, Figure 2). In addition, ST4 activity during detection level 1 was lower in participants who reported higher resilience (R = -0.41, p = 0.026, Figure 3).

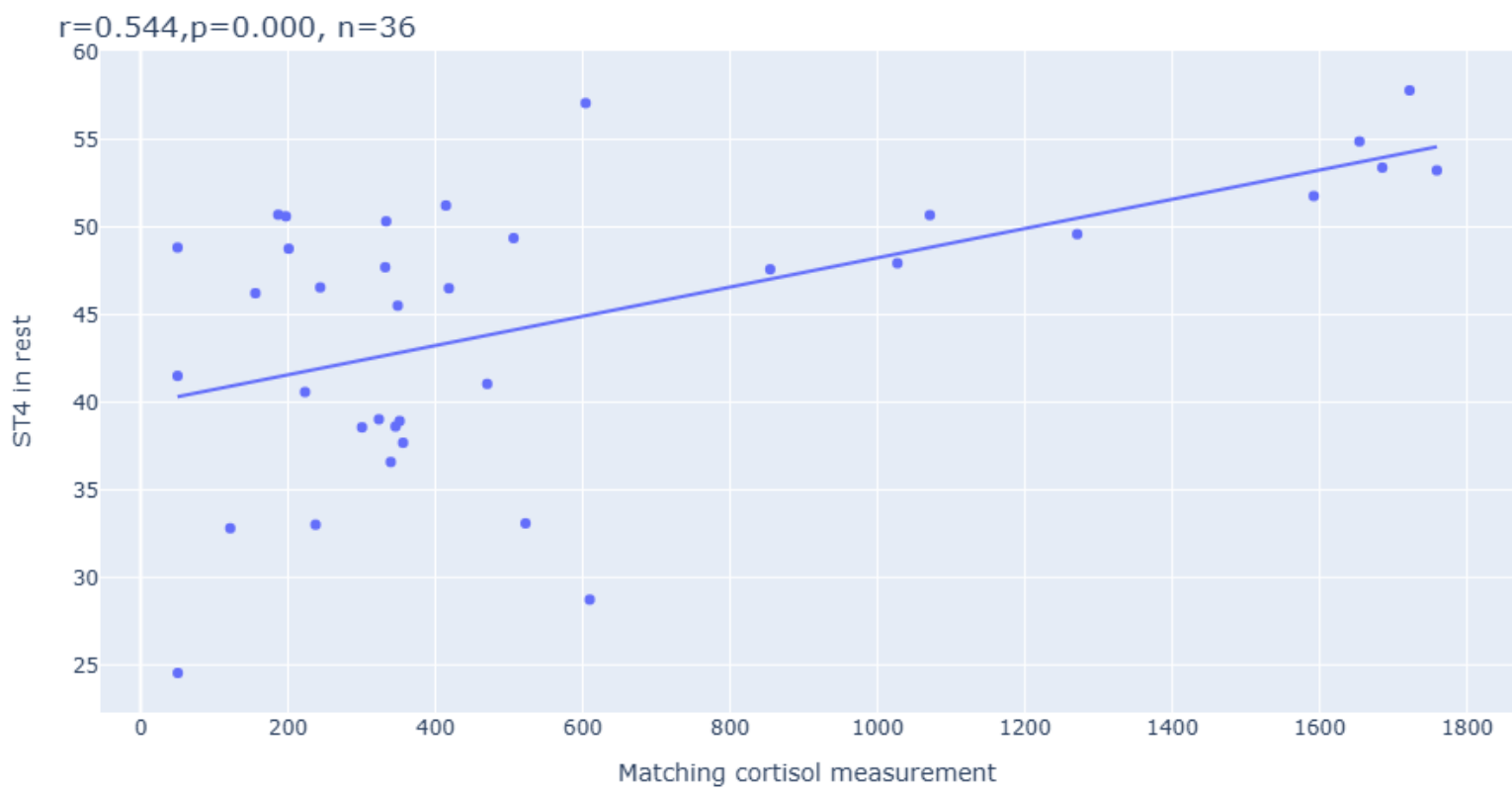

**Figure 2:** Correlation between ST4 values on the y-axis and cortisol levels on the x-axis for participants in Study 1. Each point represents an individual participant. The regression line indicates the direction and strength of the relationship.

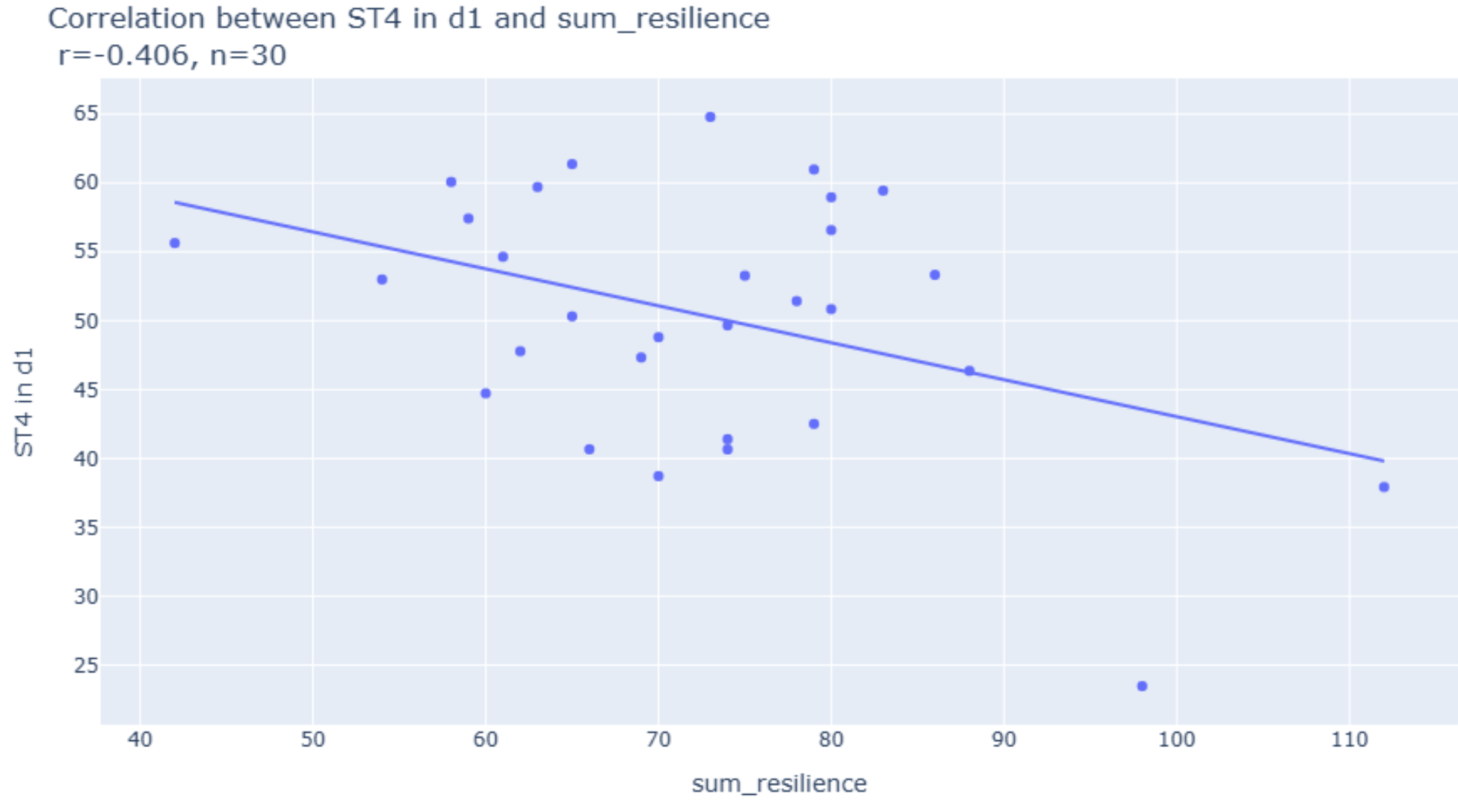

**Figure 3.** Correlation between ST4 values on the y-axis and subjective resilience levels on the x-axis for participants in Study 1. Each point represents an individual participant. The regression line indicates the direction and strength of the relationship.

For T2, greater differentiation between high mental load condition (3-back) and resting state, was associated with higher cortisol levels (R = 0.32, p = 0.038, Figure 4).

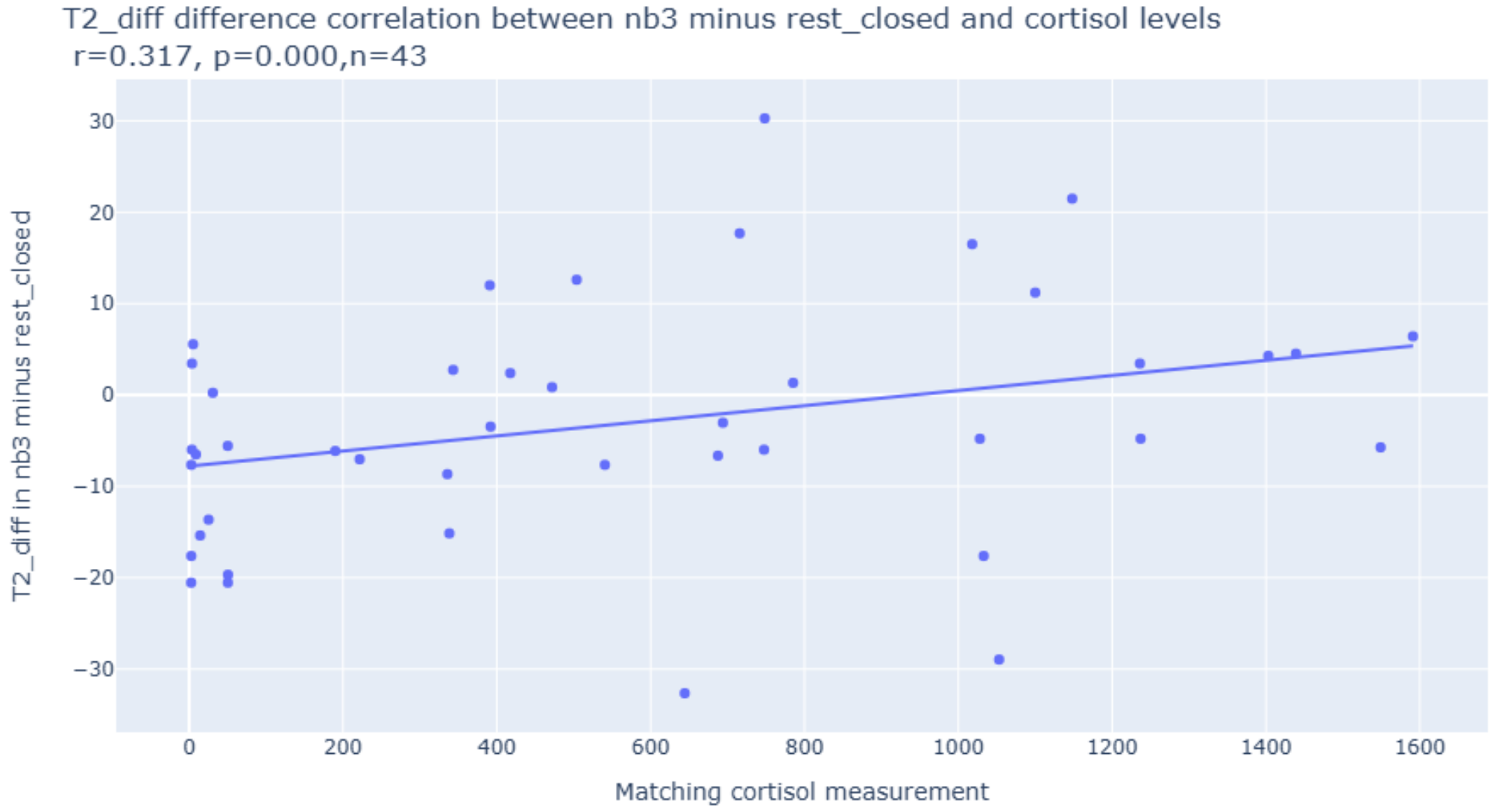

**Figure 4.** Correlation between T2 difference between 3-back and resting state on the y-axis and cortisol levels on the x-axis for participants in Study 1. Each point represents an individual participant. The regression line indicates the direction and strength of the relationship.

### 3.2 Study 2

ST4 activity during the stressful task of preparing for the job interview was lower in participants with higher HRV values of RMSSD (R = -0.24, p = 0.028, Figure 5) and during the recovery (R = -0.32, p = 0.04, Figure 6). In addition, ST4 showed negative correlations with SDNN during recovery and cognitive battery (R = -0.41, p = 0.01, Figure 7, and R = -0.47, p = 0.003, Figure 8, respectively).

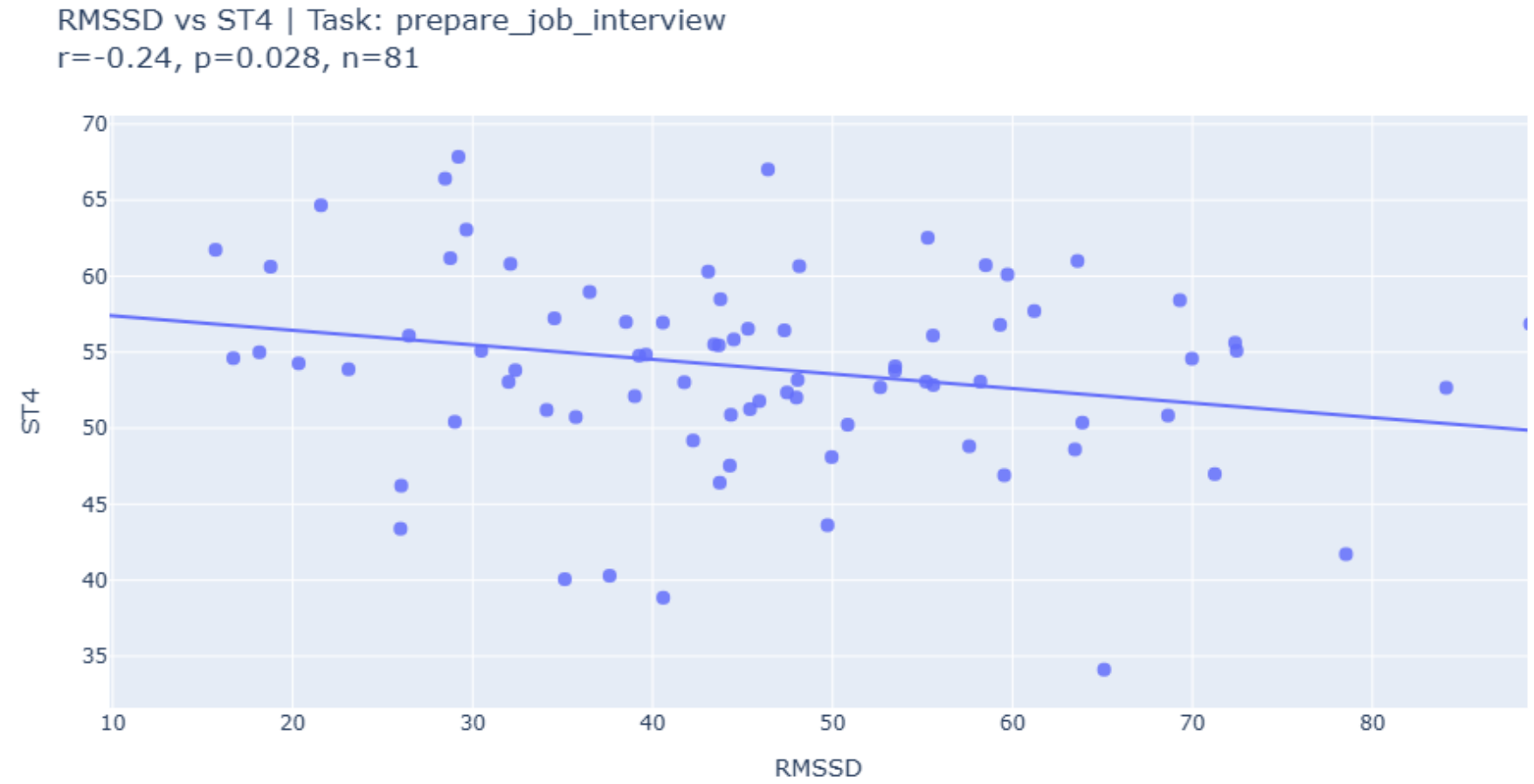

**Figure 5.** Correlation between ST4 values during preparation for job interview on the y-axis and RMSSD values on the x-axis for participants in Study 2. Each point represents an individual participant. The regression line indicates the direction and strength of the relationship.

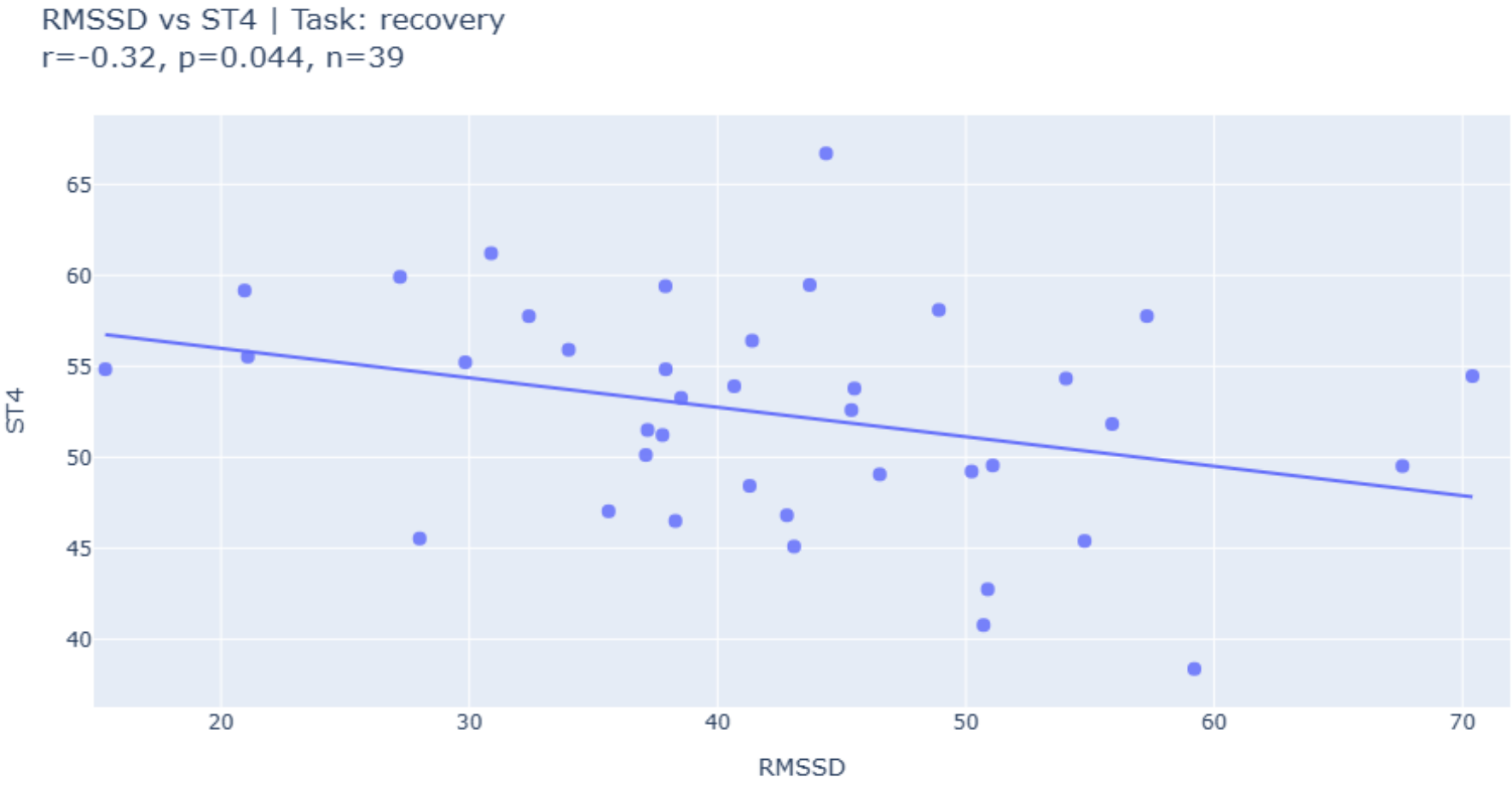

**Figure 6.** Correlation between RMSSD values on the x-axis and ST4 values during recovery on the y-axis for participants in Study 2. Each point represents an individual participant. The regression line indicates the direction and strength of the relationship.

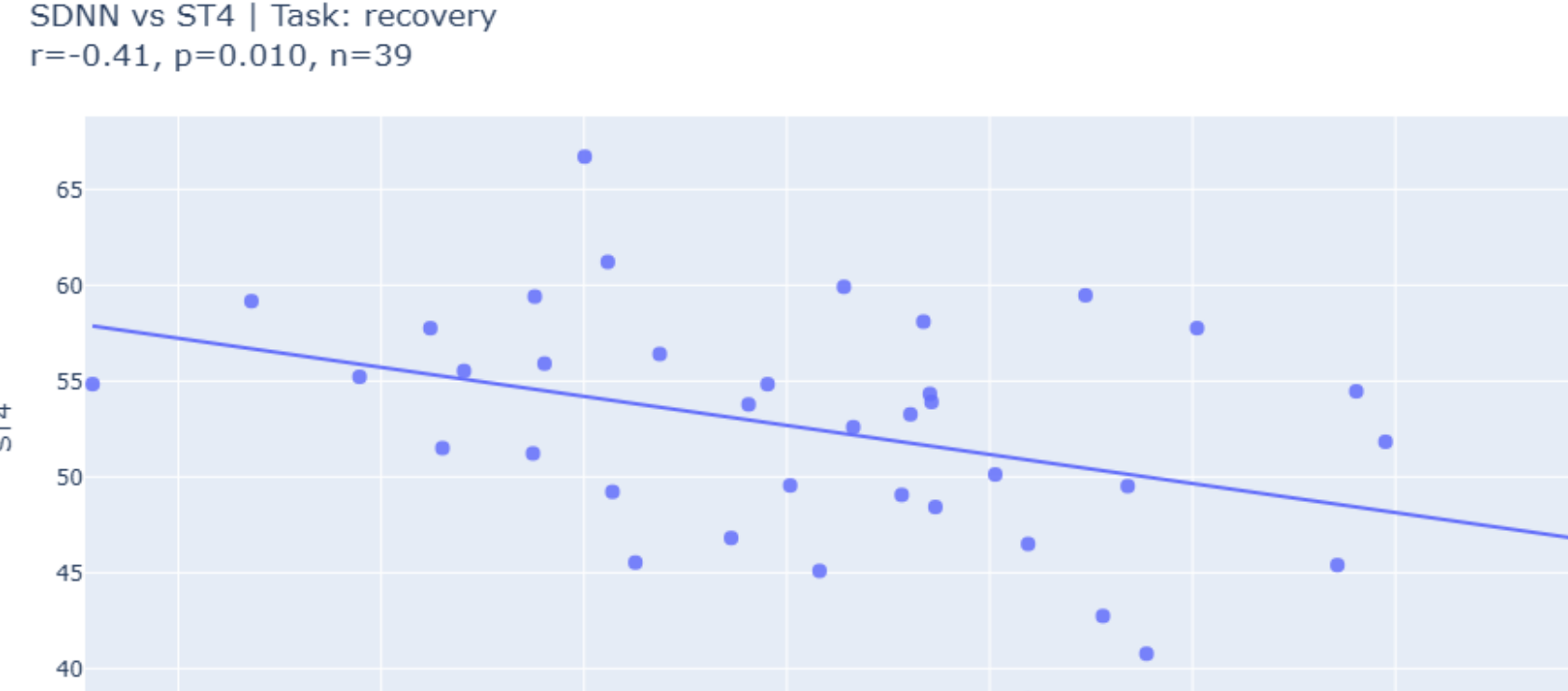

**Figure 7.** Correlation between SDNN values on the x-axis and ST4 values during recovery on the y-axis for participants in Study 2. Each point represents an individual participant. The regression line indicates the direction and strength of the relationship.

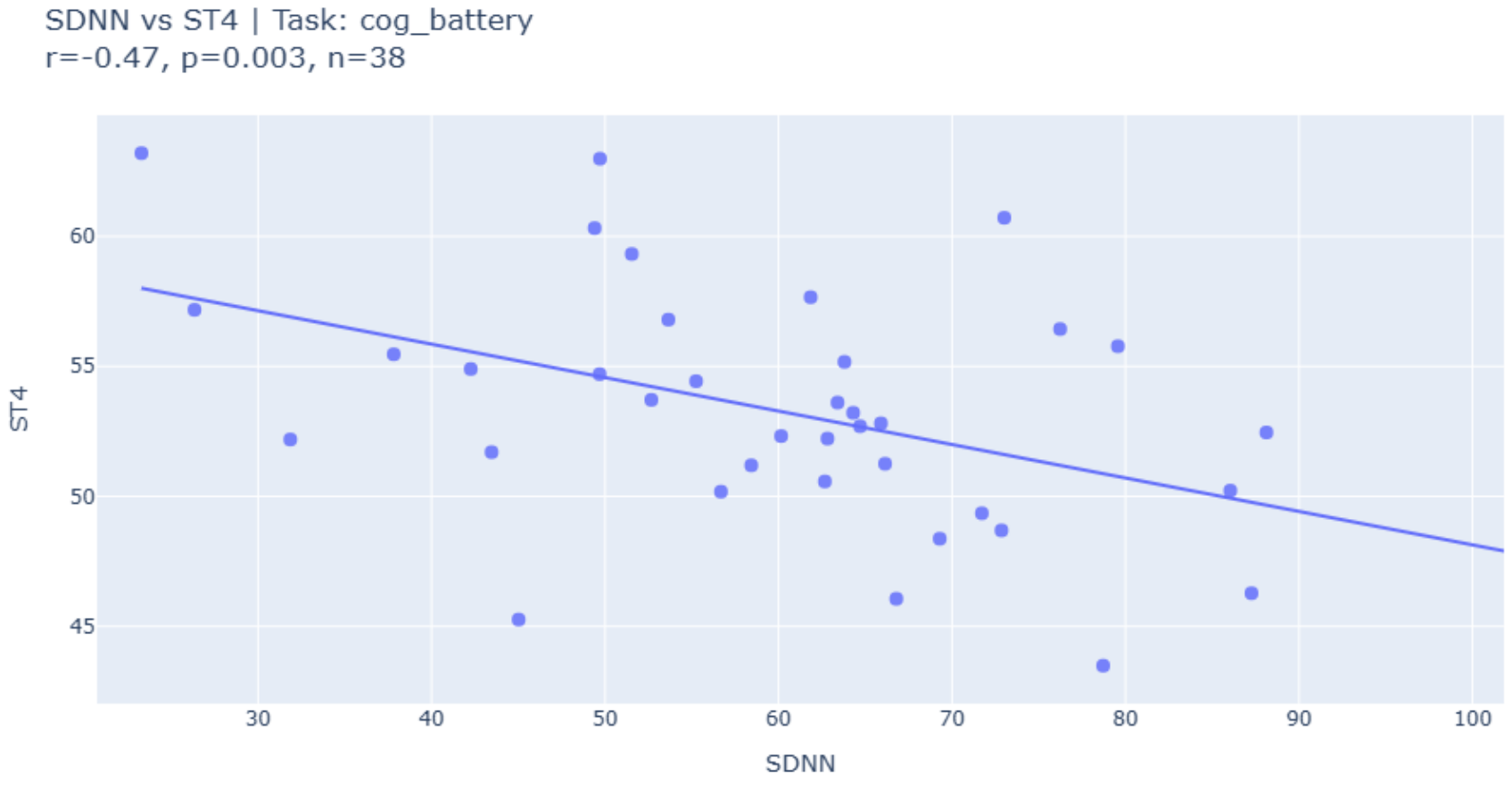

**Figure 8.** Correlation between ST4 values during preparation to job interview on the y-axis and RMSSD values on the x-axis for participants in Study 2. Each point represents an individual participant. The regression line indicates the direction and strength of the relationship.

T2 levels during the job interview were higher in participants with higher trait anxiety scores on the STAI ($R = 0.367$, $p = 0.026$, Figure 9). T2 was negatively correlated with SDNN during stress tasks and emotional tasks, ($R = -0.33$, $p = 0.029$, Figure 10; $R = -0.35$, $p = 0.021$, Figure 11). Finally, T2 was negatively correlated with RMSSD during emotional tasks ($R = -0.32$, $p = 0.036$, Figure 12).

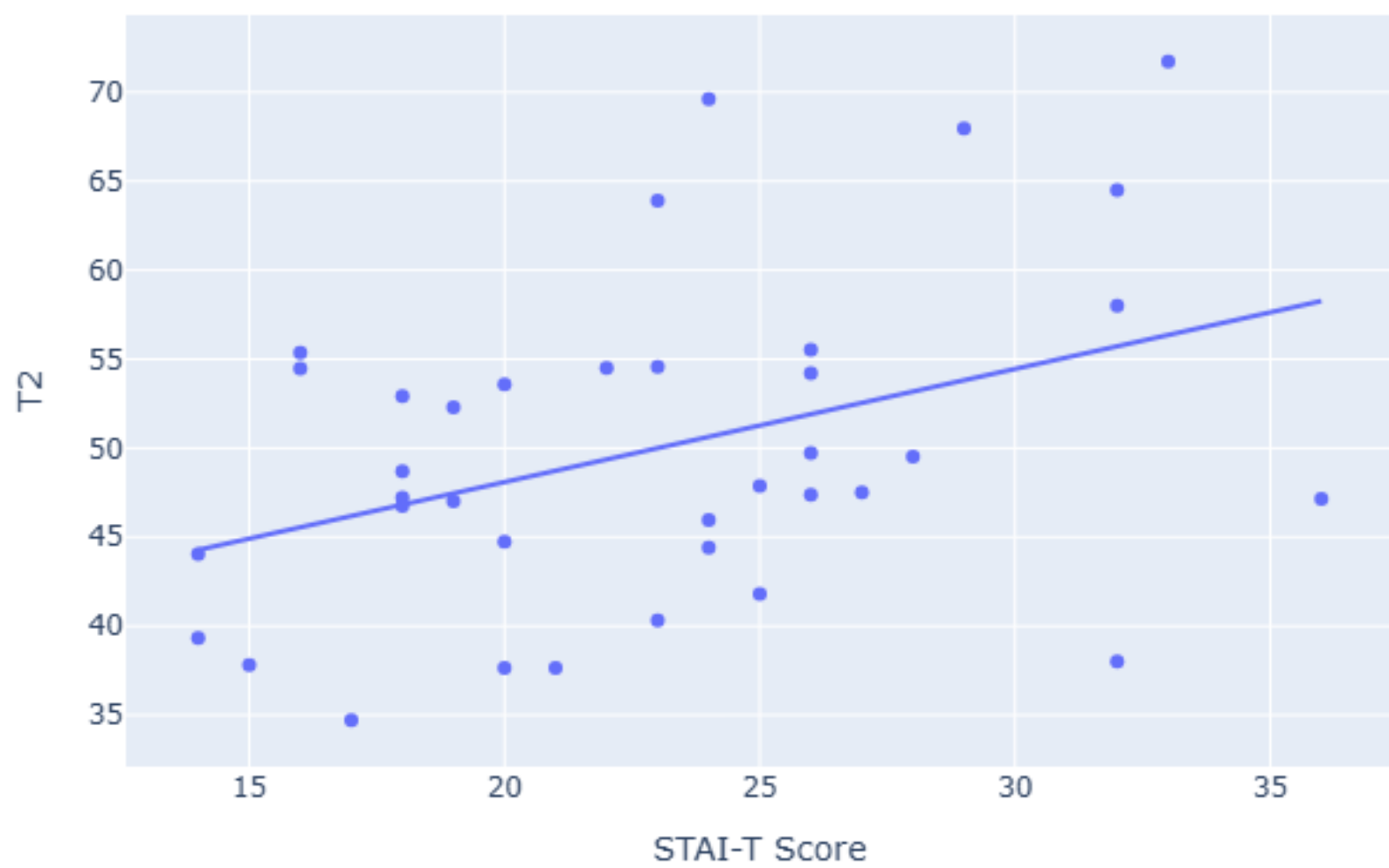

**Figure 9.** Correlation between T2 values during the job interview on the y-axis and Trait anxiety values on the x-axis for participants in Study 2. Each point represents an individual participant. The regression line indicates the direction and strength of the relationship.

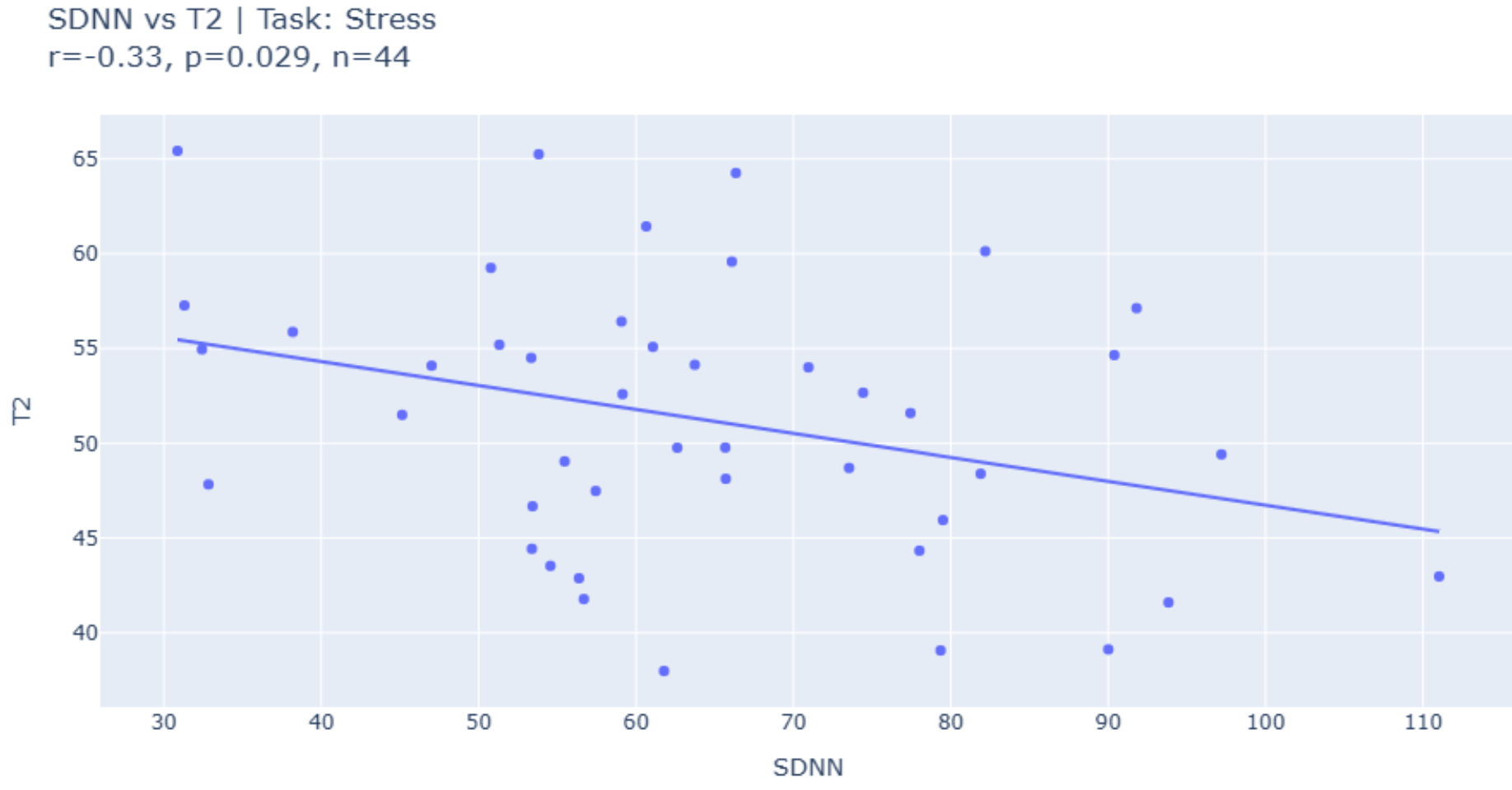

**Figure 10.** Correlation between SDNN values on the x-axis and T2 values during stressful tasks on y-axis for participants in Study 2. Each point represents an individual participant. The regression line indicates the direction and strength of the relationship.

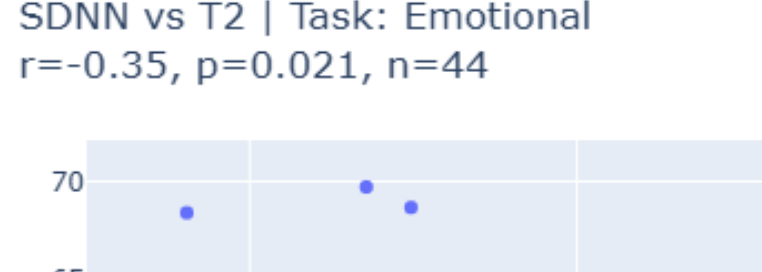

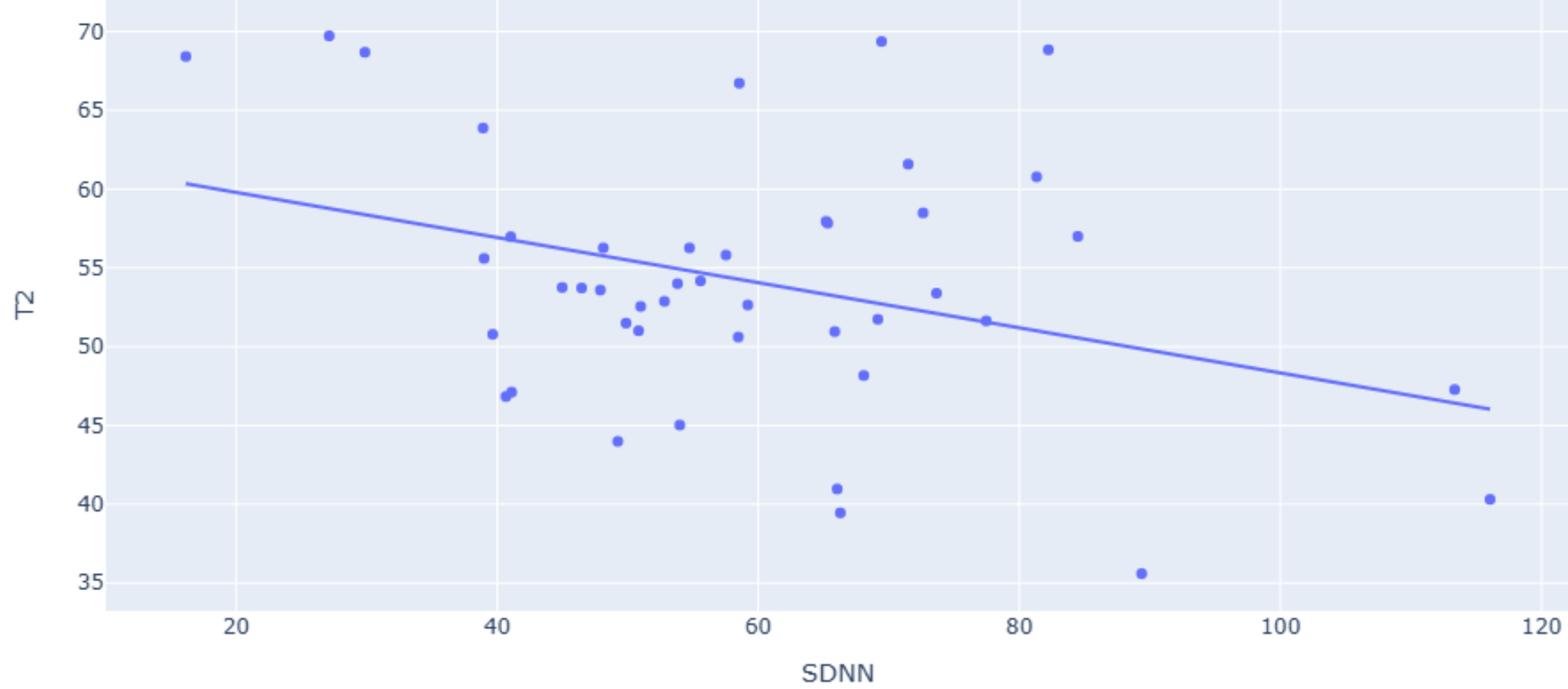

**Figure 11.** Correlation between SDNN values on the x-axis and T2 values during emotional tasks on y-axis for participants in Study 2. Each point represents an individual participant. The regression line indicates the direction and strength of the relationship.

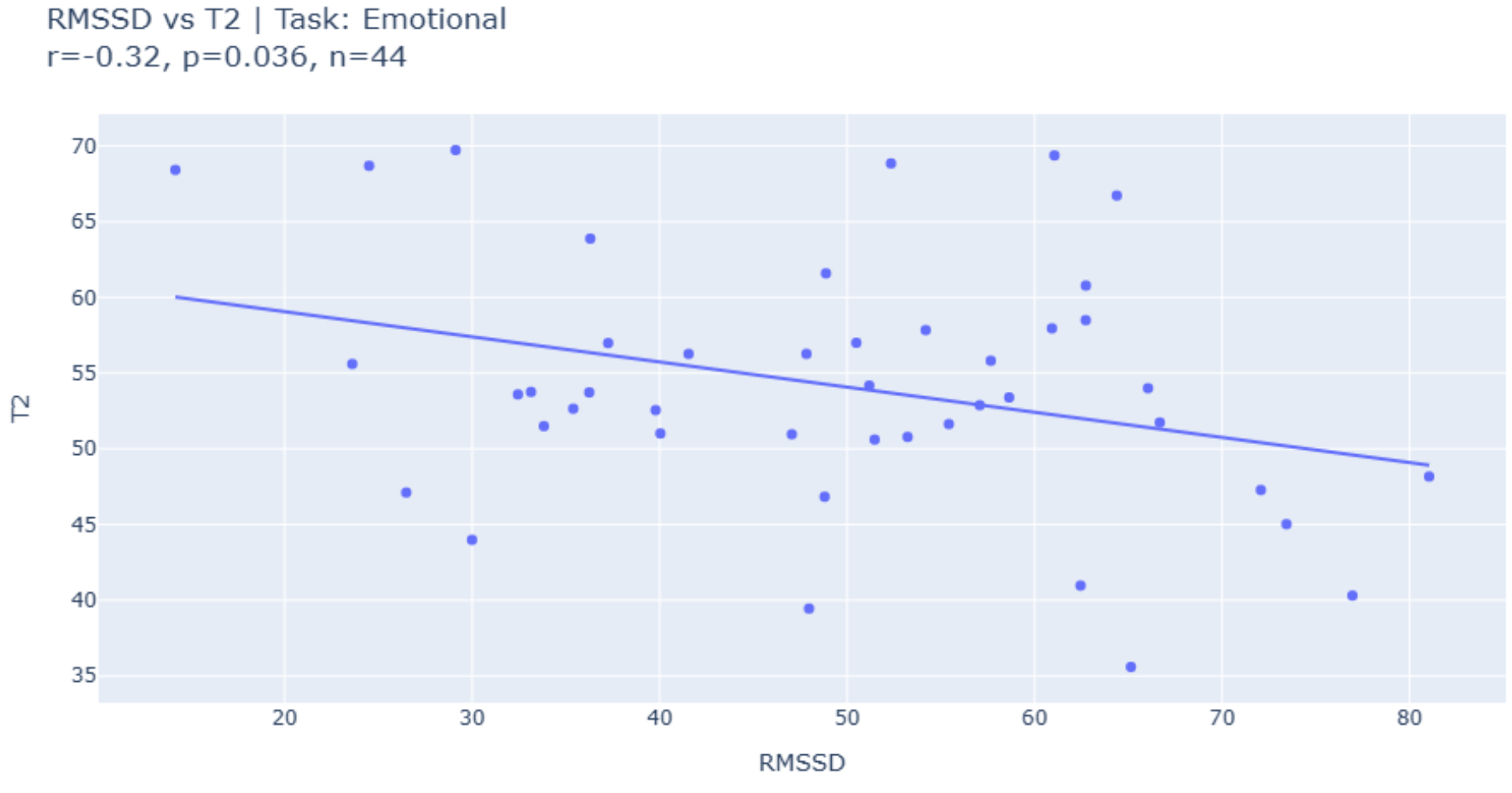

**Figure 12.** Correlation between RMSSD values on the x-axis and T2 values during emotional tasks on y-axis for participants in Study 2. Each point represents an individual participant. The regression line indicates the direction and strength of the relationship.

## 4. Discussion

The present research investigated associations between two Neurosteer-derived EEG biomarkers, ST4 and T2, and multiple validated measures of stress, including salivary cortisol, heart rate variability (HRV), pulse pressure, and subjective self-reports. Results across the two large-scale studies revealed consistent and distinct correlational patterns for the two biomarkers, extending prior evidence that EEG-derived features can serve as individualized indices of stress reactivity [4,11,12,16].

In Study 1, ST4 levels during resting state positively correlated with cortisol concentrations, suggesting sensitivity to baseline physiological stress levels. Cortisol has long been considered a robust biomarker of HPA axis activation and chronic stress burden [1,3,5], and the alignment of ST4 with cortisol supports its validity as a neural index of physiological arousal. ST4 also showed negative associations with resilience during low-load detection, consistent with the notion that adaptive coping strategies buffer against neural hyperactivation under challenge [2,17]. For T2, greater differentiation between high mental load (3-back) and rest was associated with elevated cortisol, again implicating HPA axis activation [5]. T2 was lower during positive and negative lexical trials among participants with higher resilience, and it correlated positively with heart rate, indicating sensitivity to sympathetic arousal. In Study 2, ST4 was negatively associated with HRV indices (RMSSD during job-interview preparation and recovery; SDNN during recovery and the cognitive battery), reinforcing the view that reduced autonomic flexibility co-occurs with heightened neural strain [7]. By contrast, T2 was negatively correlated with HRV (SDNN during stress and emotional tasks; RMSSD during emotional tasks) and was higher among participants with greater trait anxiety, indicating sensitivity to emotional-reactive regulation under stress [6,7,8,9,10].

However, several limitations should be noted. Although the analyses were exploratory, we applied a conservative threshold of $r \geq 0.30$ to interpret correlations as meaningful, providing some control over false discovery while prioritizing effect sizes [26]; nonetheless, replication with larger samples and formal multiple-comparisons correction is needed to confirm robustness [11,13]. Our samples consisted of healthy adults, which may limit generalizability to clinical groups with chronic stress, anxiety disorders, or neurodegenerative conditions [1,25]. Although multimodal measures such as cortisol and HRV provided convergent validation, both can be influenced by circadian, situational, or lifestyle factors not fully controlled here [3,5,6]. Finally, while single-channel hdrEEG offers ecological feasibility, its reduced spatial resolution compared to multichannel EEG may limit source localization of stress-related activity [11,12,16]. Future work should therefore combine mobile EEG with longitudinal designs and broader clinical sampling to establish ST4 and T2 as reliable, individualized biomarkers of stress regulation and vulnerability [4,17].

T2 demonstrated a different but complementary profile. In Study 1, greater differentiation of T2 between positive and negative lexical items, as well as between high mental load and resting state, was associated with elevated cortisol, again implicating HPA axis activation [5]. At the same time, lower T2 activity during valenced word processing was associated with higher resilience, suggesting that T2 reflects vulnerability to emotional interference under stress. This aligns with prior ERP work showing valence-related modulation of word processing in stress-sensitive populations [14]. T2 also correlated positively with heart rate in Study 1, consistent with sympathetic arousal. In Study 2, however, T2 correlated negatively with HRV measures (RMSSD and SDNN) across stress-inducing and emotional tasks, and it was positively associated with trait anxiety on the STAI, highlighting T2 as a biomarker of emotional-reactive regulation that co-varies with reduced autonomic flexibility during affectively demanding contexts [6,7,8,9,10].

Taken together, these findings suggest that ST4 and T2 represent complementary dimensions of stress processing. ST4 primarily indexes stress arousal and cognitive load-related strain, with strong links to cortisol and reduced HRV, situating it as a neural correlate of sympathetic dominance and HPA axis activation [1,4,6]. In contrast, T2 is more strongly modulated by emotional and autonomic processes, linking EEG activity to resilience and trait anxiety, and showing sensitivity to reduced HRV under emotional and stress challenges [14,9,7].

By combining findings from both studies, a clear pattern emerges: ST4 captures the physiological arousal dimension of stress, while T2 captures emotional-reactive regulation. Together, they provide a multidimensional picture of stress that cannot be fully captured by self-report, cortisol, or HRV measures alone [4,11,12]. This dual-biomarker approach highlights the potential of hdrEEG to disentangle overlapping components of the stress response: physiological strain reflected in ST4, versus emotional and regulatory processes reflected in T2. Such differentiation is particularly important given individual variability in stress reactivity, where some individuals show elevated physiological strain with minimal subjective distress, while others exhibit strong emotional reactivity despite preserved physiological adaptation [3,6,2].

This work directly addresses the research gap outlined in the introduction. Previous EEG stress research has largely emphasized group-level effects or acute laboratory manipulations [11,13,12], with limited focus on individual-level biomarkers. The present studies demonstrate that single-channel hdrEEG can yield personalized biomarkers of stress that correlate with gold-standard physiological measures (cortisol, HRV, blood pressure) and validated subjective assessments (STAI, resilience, burnout). By validating ST4 and T2 against independent multimodal indices, these findings extend prior work on single-channel EEG [18,16] and show that portable neural recordings can move beyond cognitive load detection [15] to serve as individualized stress assessment tools.

Importantly, the integration of emotional tasks such as music and lexical decisions alongside cognitive and stress-inducing paradigms provided a richer assessment of stress reactivity across domains. Emotional tasks are known to reliably elicit affective responses [14,24], and their inclusion allowed the demonstration that T2, in particular, is sensitive to affective and autonomic regulation. The divergence of ST4 and T2 profiles underscores that stress is not a unitary construct but a multifaceted process with separable neural signatures. This multidimensional characterization has critical implications for both clinical screening (e.g., early detection of stress vulnerability linked to burnout, anxiety, or cognitive decline [1,25]) and occupational monitoring (e.g., real-time stress regulation assessment in high-demand environments [4,12,16]).

Taken together, these findings indicate that ST4 and T2 represent distinct yet complementary dimensions of the human stress response. ST4 reflects physiological arousal and cognitive strain, linking directly to cortisol, cardiovascular load, and reduced HRV, while T2 captures emotional and autonomic regulation, showing sensitivity to resilience, emotional interference, and trait anxiety. Together, they offer a more complete and personalized view of stress than any single measure, pointing toward a promising approach for individual-level stress detection in both clinical and real-world settings.

## 5. References

1. Chrousos, G. P. (2009). Stress and disorders of the stress system. *Nature Reviews Endocrinology*, 5(7), 374-381.
2. Yaribeygi, H., Panahi, Y., Sahraei, H., Johnston, T. P., & Sahebkar, A. (2017). The impact of stress on body function: A review. *EXCLI Journal*, 16, 1057-1072.
3. Dickerson, S. S., & Kemeny, M. E. (2004). Acute stressors and cortisol responses: A theoretical integration and synthesis of laboratory research. *Psychological Bulletin*, 130(3), 355-391.
4. Giannakakis, G., Grigoriadis, D., Giannakaki, K., Simantiraki, O., Roniotis, A., & Tsiknakis, M. (2022). Review on psychological stress detection using biosignals. *IEEE Transactions on Affective Computing*, 13(1), 440-460.
5. Hellhammer, D. H., Wüst, S., & Kudielka, B. M. (2009). Salivary cortisol as a biomarker in stress research. *Psychoneuroendocrinology*, 34(2), 163-171.
6. Kim, H. G., Cheon, E. J., Bai, D. S., Lee, Y. H., & Koo, B. H. (2018). Stress and heart rate variability: A meta-analysis and review of the literature. *Psychiatry Investigation*, 15(3), 235-245.
7. Thayer, J. F., Åhs, F., Fredrikson, M., Sollers III, J. J., & Wager, T. D. (2012). A meta-analysis of heart rate variability and neuroimaging studies: Implications for heart rate variability as a marker of stress and health. *Neuroscience & Biobehavioral Reviews*, 36(2), 747-756.
8. Yugar, L. B. T., Yugar-Toledo, J. C., Dinamarco, N., Sedenho-Prado, L. G., Moreno, B. V. D., Rubio, T. D. A., ... & Moreno, H. (2023). The role of heart rate variability (HRV) in different hypertensive syndromes. *Diagnostics*, *13*(4), 785.
9. Spielberger, C. D., Gorsuch, R. L., Lushene, R., Vagg, P. R., & Jacobs, G. A. (1983). *Manual for the State-Trait Anxiety Inventory*. Consulting Psychologists Press.
10. Valente, G., Diotaiuti, P., Corrado, S., Tosti, B., Zanon, A., & Mancone, S. (2025). Validity and measurement invariance of abbreviated scales of the State-Trait Anxiety Inventory (STAI-Y) in a population of Italian young adults. *Frontiers in Psychology*, *16*, 1443375.
11. Katmah, R., Al-Shargie, F., Tariq, U., Babiloni, F., Al-Nashash, H., & Yahya, N. (2021). A review on mental stress assessment methods using EEG signals. *Sensors*, 21(15), 5043.
12. Shafiq, A., Afzal, A., Mahmood, M. T., & Adeel, A. (2022). Wearable EEG-based brain-computer interface for stress monitoring. *Sensors*, 22(19), 7350.
13. Pei, D., Tirumala, S., Tun, K. T., Ajendla, A., & Vinjamuri, R. (2024). Identifying neurophysiological correlates of stress. *Frontiers in Medical Engineering*, 2, 1434753.
14. Imbir, K. K., Spustek, T., & Żygierewicz, J. (2016). Effects of valence and origin of emotions in word processing evidenced by event related potential correlates in a lexical decision task. *Frontiers in Psychology*, 7, 271.

15. Salai, M., Vasserman, G., & Levy, I. (2016). A method for automatic detection of mental load in multimedia content. *Multimedia Tools and Applications*, 75(22), 15001-15022.
16. Vos, G., Ebrahimpour, M., van Eijk, L., Sarnyai, Z., & Azghadi, M. R. (2025). Stress monitoring using low-cost electroencephalogram devices: A systematic literature review. *International Journal of Medical Informatics*, 185, 105397.
17. Maimon, N. B., Molcho, L., Zaimer, T., Chibotero, O., Intrator, N., & Yahalom, E. (2025). Dissociating cognitive load and stress responses using single-channel EEG: Behavioral and neural correlates of anxiety across cognitive states. *arXiv preprint arXiv:2507.10093*.
18. Molcho, L., Maimon, N. B., Regev-Plotnik, N., Rabinowicz, S., Intrator, N., & Sasson, A. (2022). Single-channel EEG features reveal an association with cognitive decline in seniors performing auditory cognitive assessment. *Frontiers in Aging Neuroscience*, 14, 773692.
19. Molcho, L., Maimon, N.B., Zeimer, T. et al. Evaluating cognitive decline detection in aging populations with single-channel EEG features based on two studies and meta-analysis. *Sci Rep* 15, 25503 (2025).
20. Maimon, N. B., Bez, M., Drobot, D., Molcho, L., Intrator, N., Kakiashvilli, E., & Bickel, A. (2022a). Continuous monitoring of mental load during virtual simulator training for laparoscopic surgery reflects laparoscopic dexterity: A comparative study using a novel wireless device. *Frontiers in Neuroscience*, 15, 694010.
21. Garnefski, N., & Kraaij, V. (2007). The cognitive emotion regulation questionnaire. *European journal of psychological assessment*, *23*(3), 141-149.
22. Schaufeli, W. B., Bakker, A. B., Hoogduin, K., Schaap, C., & Kladler, A. (2001). On the clinical validity of the Maslach Burnout Inventory and the Burnout Measure. *Psychology & health*, *16*(5), 565-582.
23. Gomes, A. R., & Teixeira, P. M. (2016). Stress, cognitive appraisal and psychological health: Testing instruments for health professionals. *Stress and Health*, *32*(2), 167-172.
24. Weth, K., Raab, M. H., & Carbon, C. C. (2015). Investigating emotional responses to self-selected sad music via self-report and automated facial analysis. *Musicae Scientiae*, *19*(4), 412-432.
25. Walker, E. R., McGee, R. E., & Druss, B. G. (2015). Mortality in mental disorders and global disease burden implications: A systematic review and meta-analysis. *JAMA Psychiatry*, 72(4), 334-341.
26. Hemphill, J. F. (2003). Interpreting the magnitudes of correlation coefficients.